 \documentstyle[preprint,prb,aps]{revtex}
 \begin{document}
 \draft
 \title{Chiral electromagnetic waves at the boundary of optical isomers: Quantum Cotton-Mouton effect}

 \author{L. E. Zhukov$^{1}$ and M. E. Raikh$^{2}$}
 \address{$^{1}$Department of Computer Science, University of Utah, 
 Salt Lake City, Utah  84112\\
 $^{2}$Department of Physics, University of Utah, Salt Lake City, Utah 84112}
 \maketitle
 \begin{abstract}
We demonstrate that the boundary of two optical isomers with opposite
directions of the gyration vectors (both parallel to boundary) can support propagation of electromagnetic wave in the direction perpendicular to the gyration axes 
(Cotton-Mouton geometry). The components of electromagnetic field 
in this wave decay exponentially into both media. The  characteristic
decay length is of the order of the Faraday rotation length for
the propagation along the gyration axis. The remarkable property
of the boundary wave is its {\em chirality}. Namely, the wave can
propagate only in {\em one} direction determined by the relative
sign of non-diagonal components of the dielectric tensor in contacting
media. We find the dispersion law of the boundary wave for the
cases of abrupt and smooth boundaries. We also study the effect
of asymmetry between the contacting media on the boundary wave and
 generalize the results
to the case of two parallel boundaries. Finally we consider
the arrangement when the
boundaries form a random network. We argue that at a point,
 when this network
percolates,  the corresponding 
boundary waves
undergo quantum delocalization transition, similar to the quantum
Hall transition.
 \end{abstract}

 \narrowtext

 Optical properties of gyrotropic and optically active media are well known and described in
  many textbooks (see e.g. Ref. \onlinecite{Fowles}). A homogeneous gyrotropic medium is characterized by the 
 following relation between the displacement vector and electric field
 ${\bf D} = \hat\varepsilon {\bf E}$, where the tensor $\hat \varepsilon$
 has the following form

\begin{equation}
 \hat \varepsilon = \left( 
 \begin{array}{ccc}
 \varepsilon_0 & ig & 0 \\ 
 -ig & \varepsilon_0 & 0 \\
 0 & 0 &\varepsilon_1
 \end{array}
 \right). 
 \label{eq1}
 \end{equation}

 For a wave propagating along the $z$-direction,  the wave equation
 \begin{equation} 
 \nabla ( \nabla {\bf E}) - \nabla^2 {\bf E} = - \frac {1} {c^2} \frac
 {{\partial}^2 {\bf D}}{\partial t^2} 
 \label{eq2}
 \end{equation}
 has two circular polarized solutions characterized by the dispersion laws
 \begin{equation}
 k_{\pm} =\frac {\omega}{c} \sqrt {\varepsilon_0 \pm g}.
 \label{eq3}
 \end{equation}
 A linear polarized wave, incident on this medium along the axis of gyration, $z$, splits into
  two circular polarized waves propagating with velocities 
 $c/(\varepsilon_0 \pm g)^{1/2}$. Since $g$ is always much smaller than $\varepsilon_0$, this  can be viewed as
 rotation of the plane of polarization (Faraday effect).
  The distance at which the plane of polarization is rotated by $90^{\circ}$ is equal to
  $l_{\omega} =  \pi{\sqrt \varepsilon_0}c/g\omega$, where $\omega$ is the frequency of the wave.
 The direction of rotation is determined by 
 the sign of the nondiagonal component, $g$, of the tensor (\ref{eq1}). Correspondingly, in
the case of optical activity, there exist 
left-rotating ({\em levorotatory}) and right-rotating ({\em dextrorotatory}) media. Two modifications of a crystal differing by the sign of $g$ are called optical isomers. For example, the parameters of tensor $\hat \varepsilon$ for the quartz crystal
 are known to be \cite{Fowles} $\varepsilon_0 = 2.3839$, $\varepsilon_1 = 2.4118$, $g = 1.1\cdot 10^{-4}$.
 Consequently, $l_{\omega} \approx 4mm$ for yellow light $(\lambda = 589 nm)$.
Isomers with tensor $\hat\varepsilon$ in the form (1) can only belong to the
certain crystalline symmetry classes listed e.g. in Ref. \onlinecite{Landau2}.

   When the direction of propagation is perpendicular to the gyration axis, 
 there  also exist two types of solutions of the wave equation (\ref{eq2}).
 Firstly, there is a trivial solution for which only 
$E_z$ component is nonzero and 
 the spectrum is  $\omega = kc/{\sqrt \varepsilon_1}$. The nontrivial solution (Cotton-Mouton effect) corresponds to 
 polarization perpendicular to the axis of gyration
 \begin{equation}
 {\bf E} = E_0(1,\frac{c^2 k_y^2-\omega^2\varepsilon_0}{c^2 k_xk_y+ig\omega^2},0)e^{ik_xx + ik_yy},
 \end{equation}
 with the dispersion law
 \begin{equation}
 \label{cotton}
 k_x^2 + k_y^2 = \frac {\omega^2}{c^2} (\varepsilon_0-\frac{g^2}{\varepsilon_0}).
 \end{equation}

 The phenomenon of surface electromagnetic waves is also a well-studied subject (see e.g. Ref. \onlinecite{Yariv}). 
 Surface waves
 can propagate along  the boundary of two isotropic media with dielectric constants $\varepsilon_a(\omega)$ and $\varepsilon_b(\omega)$
 when two conditions are met \cite{Yariv}: ({\it i}) $\varepsilon_a(\omega) \cdot\varepsilon_b(\omega) <0$ and ({\it ii}) $\varepsilon_a(\omega) + \varepsilon_b(\omega) <0$. The polarization of these waves is TM (transverse-magnetic). 
For such waves
the localization of electromagnetic field at the boundary is provided by negative sign of $\varepsilon$ in one of the contacting media. Interesting examples
of surface electromagnetic wave at the boundary of two {\em transparent} 
{\em media} were given in Refs. \onlinecite{Dyakonov} and \onlinecite{Averkiev}.
 In Ref. \onlinecite{Dyakonov} the boundary between isotropic and uniaxial media is considered. In Ref. \onlinecite{Averkiev} both contacting media represent uniaxial
crystals with certain mismatch in the directions of optical axes.

 In the present paper we  show that the boundary of two optical isomers with 
 opposite directions of axes of optical activity (both parallel to the boundary) can also
 support propagation of localized
  electromagnetic waves. In contrast to the conventional surface plasmons \cite{Yariv},
 these states exist for {\it positive} values of components $\varepsilon_0$ 
and  $\varepsilon_1$.
 The distinguishing feature of these boundary waves  is a {\it unidirectional} character of  propagation, i.e. they propagate along the boundary of two media with opposite gyration vectors
{\em only} {\em in} {\em one} {\em direction} which is determined by the relative sign of $g$.

 Let the plane $x=0$ be a boundary between two isomers. For generality we consider $g$
 to be some  function of $x$ , $g(x)$, changing its value from $g(x) = g_0$ to $g(x) = -g_0$ within some region around  $x=0$. In the Cotton-Mouton 
 geometry, when only $x$ and $y$ components of electric field are nonzero, the system of
 equations for $E_x$, $E_y$ which follows from Eq. (\ref{eq2}) takes the form
  \begin{eqnarray}
 \frac {{\partial}^2 E_y}{\partial x \partial y} - \frac {{\partial}^2 E_x}{\partial y^2}  = \frac {\omega^2} {c^2}(\varepsilon_0 E_x + ig(x)E_y)\\
 \frac {{\partial}^2 E_x}{\partial x \partial y} - \frac {{\partial}^2 E_y}{\partial x^2}  = \frac {\omega^2} {c^2}(\varepsilon_0 E_y - ig(x)E_x).
 \end{eqnarray}
 Let us  search for a solution propagating along the $y$ axis: $E_x \propto e^{iky}$,
 $E_y \propto e^{iky}$. Then from Eq. (6) we can express $E_x$ through $E_y$ 
 \begin{equation}
 E_x = \frac {ik c^2\frac {{\partial} E_y}{\partial x} - i\omega^2E_yg(x)}
 {\varepsilon_0 \omega^2 - k^2 c^2}.
 \label{eq6}
 \end{equation}
 Substituting Eq. (\ref{eq6}) into Eq. (7) we get
 \begin{equation}
 - \frac {{\partial}^2 E_y}{\partial x^2} + 
 \Biggl[\frac {k} {\varepsilon_0} \frac {\partial g(x)} {\partial x} + {\frac {\omega^2} {c^2
 \varepsilon_0}} (g^2(x)-g_0^2)\Biggr]
  E_y = \Biggl[{\frac {\omega^2} {c^2} }(\varepsilon_0 - \frac{g_0^2}{\varepsilon_0}) - k^2\Biggl]E_y.
 \label{schred1}
 \end{equation}
 Note now, that Eq. (\ref{schred1}) has the form of the Schroedinger equation with an effective potential
 \begin{equation}
 U(x) = \frac {k} {\varepsilon_0} \frac {\partial g(x)} {\partial x} + {\frac {\omega^2} {c^2
 \varepsilon_0}}(g^2(x)-g_0^2),
 \end{equation}
 and energy
 \begin{equation}
 {\cal E} = \frac {\omega^2} {c^2} (\varepsilon_0 - \frac{g_0^2}{\varepsilon_0}) - k^2.
 \label{energy}
 \end{equation}

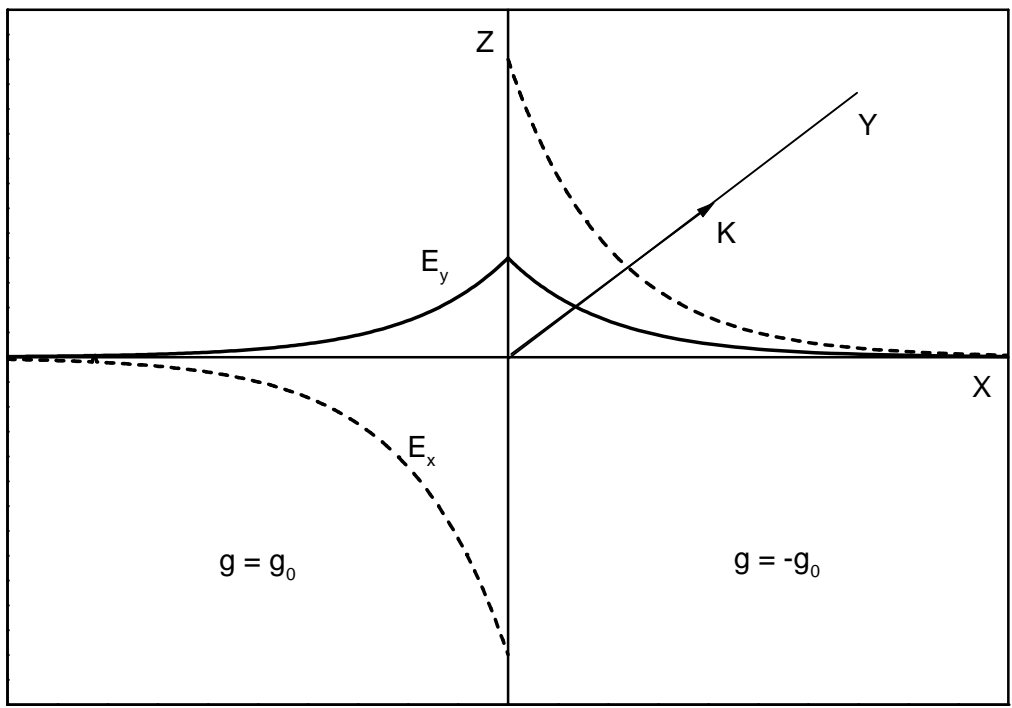
 \begin{figure}
 \caption{The boundary of two optical isomers $(x = 0)$. Boundary wave propagates 
along the $y$-axis. The decay of $E_x$ (dashed curve) and $E_y$ (solid curve) components of electric field away from the boundary $x=0$
is shown schematically.}
 \label{fig1}
 \end{figure}

 Consider first the simplest case when the transitional region between two media is 
 infinitely narrow: $g(x) = -g_0(2\theta(x) - 1)$, where $g_0>0$ and $\theta(x)$ is a step-function (see Fig.~\ref{fig1}). Then $U(x)$ takes the form $U(x)=-2g_0k\delta(x)/\varepsilon_0$. 
 It is seen that the potential  is attractive for the positive values of  $k$ and  repulsive  
 for the negative $k$. This illustrates the above statement about the {\it chirality}
 of the boundary states.
 It is well known that there exists only a single bound state 
 in attractive $\delta$-potential with an arbitrary magnitude \cite{Landau}. The corresponding 
 ``binding energy'' is equal to ${\cal E} = -g_0^2k^2/\varepsilon_0^2$. This leads to the following dispersion law for electromagnetic wave propagating along the abrupt boundary between two media with $g = \pm g_0$
 \begin{equation}
 \label{first}
 \omega = \frac {ck}{\sqrt \varepsilon_0}.
\label{dispersion}
 \end{equation}
 The origin of  the  localized  boundary  state Eq. (\ref{first}) is  the  following. 
 It is  seen from Eq.~(\ref{cotton}) that the  maximal  possible wave vector inside each of the contacting media 
 is equal to $\frac {\omega}{c}$~$(\varepsilon_0-$~$\frac{g_0^2}{\varepsilon_0})^{1/2}$. 
 The wave vector determined by Eq. (\ref{first}) exceeds this maximal wave vector. As a
 result, both components of electric field decay to the left and to the right from
 the boundary $x=0$ 
 \begin{eqnarray}
\label{fields1}
 E_y(x) &=& E_y(0)e^{-q|x|} \\
 E_x(x) &=& sign(x)\frac{i(\varepsilon_0^2+g_0^2)}{2g_0\varepsilon_0}E_y(0)e^{-q|x|} \nonumber, 
 \end{eqnarray}
 where $q = |kg_0|/\varepsilon_0$ (Fig. \ref{fig1}).
 The characteristic decay length, $1/q$, can be expressed through the length of
 rotation of plane of polarization of light in the Faraday geometry: 
 $1/q =  l_{\omega}/\pi$.

 It is easy to establish certain properties of chiral boundary states.

{\em a)} {\em Asymmetry} {\em between} {\em the}  {\em contacting}
{\em media}.
 Suppose that the mirror symmetry between the contacting media is
lifted, say, due to
 external magnetic field  applied in the $z$ direction.
This amounts to the following modification of $g(x): ~~~
 g(x) = -g_0(2\theta(x)-1) + g_1$, where $g_1$ is proportional to the magnetic field. 
 Then it is straightforward to check that the dispersion law
 Eq. (\ref{first}) remains unchanged, while the magnitude of $g_1$ remains smaller than $g_0$. When $g_1$ 
 exceeds $g_0$ the bound state {\em disappears} {\em abruptly}. Despite the dispersion law being insensitive to the asymmetry for $g_1 < g_0$, the distribution $E_y(x)$ changes strongly. For $g_1 \neq 0$ instead of 
 Eq. (\ref{fields1}) we have 
 \begin{eqnarray}
 E_y(x) &=& E_y(0)\exp\Bigl[-\frac {\omega} {c \sqrt{\varepsilon_0}} (g_0-g_1) x\Bigr], ~~~~ x>0 \\
 E_y(x) &=& E_y(0)\exp\Bigl[\frac {\omega} {c \sqrt{\varepsilon_0}}(g_0+g_1) x\Bigr], ~~~~ x<0
 \end{eqnarray}

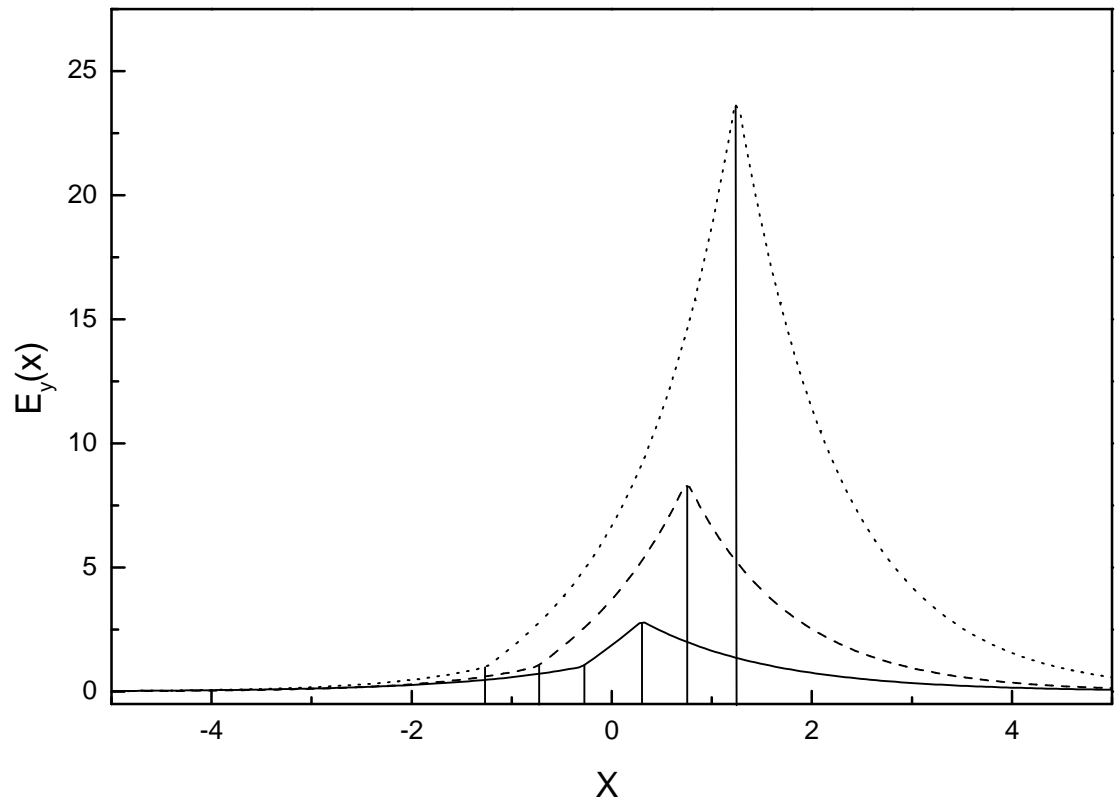
 \begin{figure}
 \caption{$E_y(x)$ component of electric field for the case of two boundaries.
 The location of the boundaries are
 $x/l_{\omega}= \pm1.25$ (dotted line);
 $x/l_{\omega}= \pm0.75$ (dashed line);
 $x/l_{\omega}= \pm0.3$ (solid line).}
 \label{fig2}
 \end{figure}

{\em b)} {\em Two} {\em boundaries}.
 Consider now the case of two boundaries
 \begin{eqnarray}
 g(x) &=& -g_0, ~~~~|x| > l/2 \\
 g(x) &=& ~g_0, ~~~~~|x| < l/2
 \end{eqnarray}
It is straightforward to analyze Eq. (\ref{schred1}) in this case. Indeed, the effective potential takes the form 
 \begin{equation}
 U(x) = 2g_0k[\delta(x-l/2) - \delta(x+l/2)]. 
 \end{equation}
The corresponding dispersion law for the boundary state becomes
 \begin{equation}
 k^2 = \frac {\omega^2}{c^2} \bigl(\varepsilon_0-\frac{g_0^2}{\varepsilon_0}\bigr) + q_l^2,
 \end{equation}
where $q_l$ is determined by the equation
 \begin{equation}
 q_l^2 \Biggl[ \frac{1}{1 - e^{-2q_ll}} - \frac {g_0^2}{\varepsilon_0^2} \Biggl] =
  \frac {\omega^2 g_0^2}{c^2 \varepsilon_0} \Biggl[ 1 - \frac {g_0^2}{\varepsilon_0^2} \Biggl].
 \label{long_0ne}
 \end{equation}
In the limit $l \rightarrow \infty$ we return to the dispersion law Eq. (\ref{dispersion}), which  for 
 positive $k$ corresponds to the wave propagating along the boundary $ x = l/2$; negative $k$ corresponds to the wave propagating along the boundary  $x = -l/2$. Inspection of  Eq. (\ref{long_0ne})
 shows that it has solution for {\em an arbitrary small } $l$. In the case when the distance between boundaries is much smaller then the localization length $l_{\omega}$ we get from
 Eq. (\ref{long_0ne})
 \begin{equation}
 q_l = \frac {l} {\pi l_{\omega}^2} << l_{\omega}^{-1}
 \end{equation}
 This means that both states (with positive and negative $k$) are ``weakly bound''. This is 
 illustrated in Fig. \ref{fig2} for different $l/l_{\omega}$ and positive $k$. For negative $k$ the distribution
 of electric field corresponds to the change $x \rightarrow -x$.

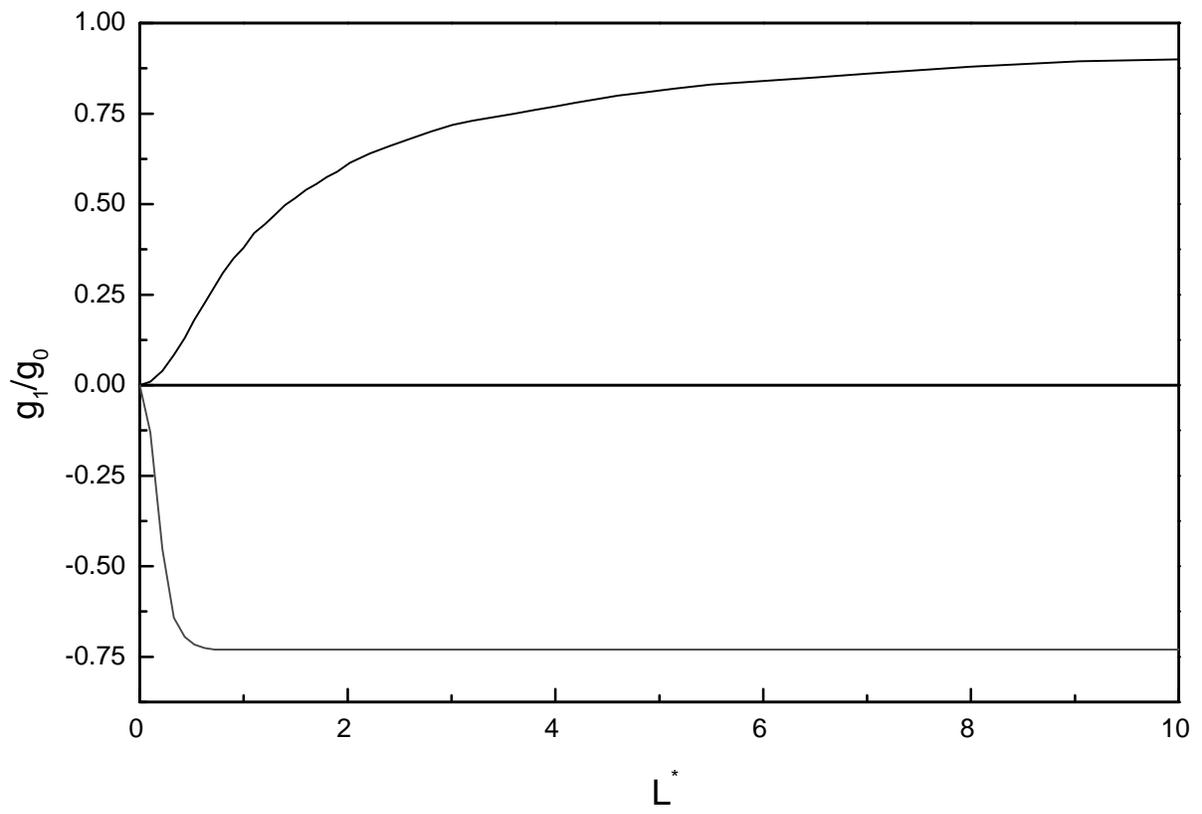
 \begin{figure}
 \caption{Critical line  $g_1/g_0$ as a function of $L^*$ for two opposite directions of
 external  magnetic field.}
 \label{fig3}
 \end{figure}

{\em c)} {\em Two} {\em boundaries} {\em with} {\em external} {\em magnetic} {\em field}.
 We have also studied the suppression of the states associated with a  pair of boundaries by an external magnetic field. In this case the critical value 
 of $g_1$ depends on the distance between boundaries. The electric field in the region
$|x|< l$ represents the superposition of exponents $\exp(\frac{\pi\kappa x}{\l_{\omega}})$ and
$\exp(-\frac{\pi\kappa x}{\l_{\omega}})$, where the dimensionless parameter $\kappa$ satisfies
the equation 
\begin{equation}
{\kappa}^2\Biggl[ 1 + \frac {2} {{\kappa}^2}  \frac {g_1}{ g_0} - \frac {2g_0^2} {\varepsilon_0^2 } +
\frac {1 + \exp(-2 L^*{\kappa})}{1-\exp(-2L^*{\kappa})} {\Biggl(1 + \frac {4 g_1}{g_0{\kappa}^2}\Biggr)}^{1/2} \Biggr] = 
2\Biggl[1 - \frac {(g_0-g_1)^2}{\varepsilon_0^2}\Biggr],
\label{kappa_eq}
\end{equation}
with $L^* = \pi l/l_{\omega}$. 
In Fig. \ref{fig3} we present the critical line of $g_1/g_0$ as a function of $L^*$, which
determines the domain where solution of Eq. (\ref{kappa_eq}) exists. Note, that the boundaries
are asymmetric with respect to the change of the direction of magnetic field (the sign of $g_1$). 

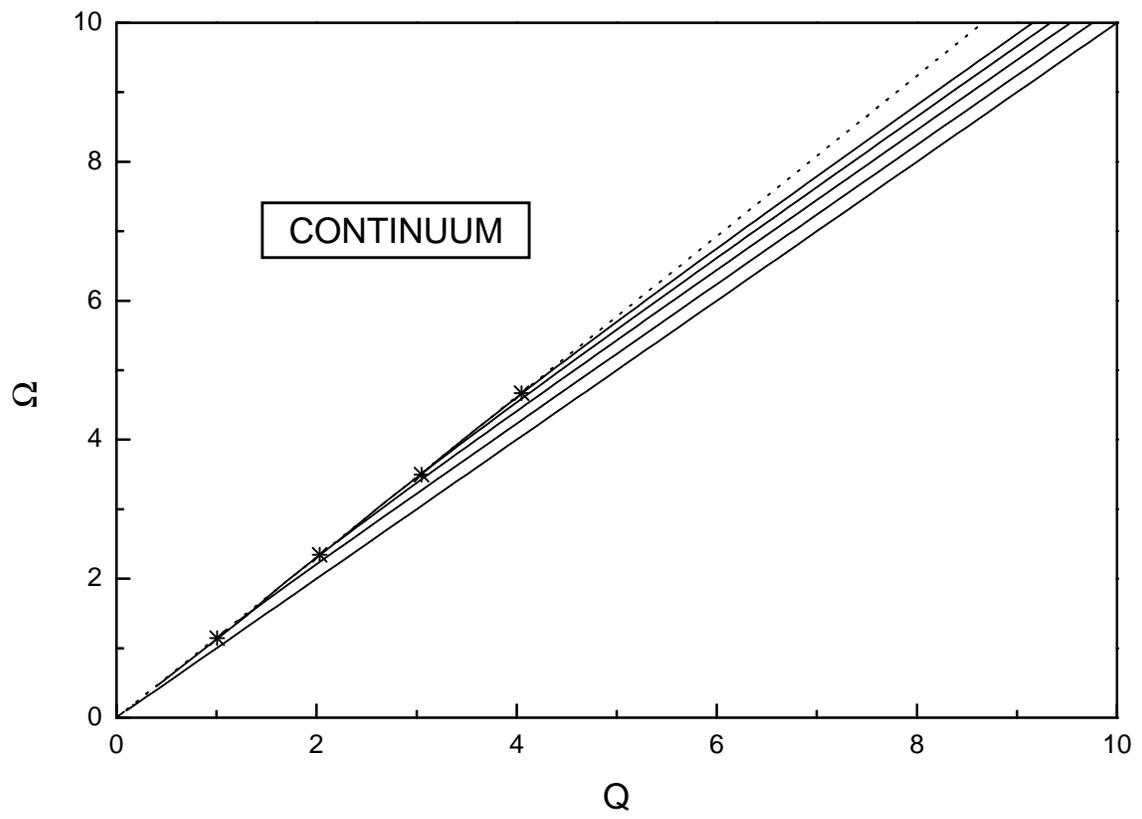
 \begin{figure}
 \caption{Dispersion 
relations for the guided modes with $n = 0,1,2,3,4$ 
are plotted using Eq.~~ (28). For illustration purposes we chose unrealistically large value of parameter $\tau = 0.5$.}
 \label{fig4}
 \end{figure}

{\em d)} {\em Smooth} {\em boundary}. 
 Next we consider the case when the boundary is smooth and has 
 a characteristic width of $b$. This, for example, can be a result of mutual diffusion
of isomers. Then $g(x)$  can be modeled by 
 $g(x) = -g_0 \tanh (x/b)$. Thus, for the effective potential $U(x)$ we get
 \begin{equation}
 \label{u}
 U(x) = - \frac { g_0 (kc^2  + \omega^2  g_0 b)} {\varepsilon_0 b c^2\cosh^2 (x/b)}.
 \end{equation}
The solutions of the Schroedinger equation with potential Eq. (\ref{u}) 
 are well known \cite{Landau} 
 \begin{equation}
 {\cal E}_n = - \frac {1}{4b^2}
 \Biggl[ {\sqrt {1+4\biggl(\frac{k g_0 b} {\varepsilon_0}+\frac {g_0^2b^2\omega^2}{\varepsilon_0 c^2}\biggl)}} - (2n+1)\Biggl]^2.
 \label{spectrum}
 \end{equation}
The mode with $n=0$ has no threshold frequency. The threshold frequencies, $\omega_n$,  for the 
 modes with $n>0$ are determined from the condition ${\cal E}_n=0$
 \begin{equation}
 \omega_n = \frac{c {\sqrt \varepsilon_0}}{2g_0b}\Biggl[{\sqrt {(2n+1)^2 - \frac {g_0^2}{\varepsilon_0^2}}} - {\sqrt { 1- \frac{g_0^2}{\varepsilon_0^2}}}\Biggr].
 \end{equation}
 The corresponding dispersion laws, $\omega_n(k)$, can be conveniently presented 
 after introducing a dimensionless frequency, $\Omega$, and the wave vector $Q$
 \begin{equation}
 \Omega= \frac {\tau \sqrt{\varepsilon_0} b} {c} \omega , ~~~~~Q = \tau b  k,
 \end{equation}
 where the dimensionless parameter $\tau$ is defined as
 \begin{equation}
 \tau = \frac {g_0} {\varepsilon_0}. 
 \end{equation}
 Then from Eqs. (\ref{energy}) and (\ref{spectrum}) we have
 \begin{equation}
 \Omega(Q) = \frac 1 2 \Biggl( \biggl[(2n+1){\tau^2} + 
 \sqrt{(2Q+1-{\tau^2})^2-4n(n+1){\tau^2}(1-{\tau^2})}\biggl]^2 - (1 + 4Q)\Biggl)^{1/2}.
 \end{equation}
The dispersion law for the first five modes 
is shown in  Fig. \ref{fig4}.

Let us consider qualitatively the situation when nondiagonal
component of the tensor $\hat \varepsilon$ is a  random function
of both coordinates $x$ and $y$. Suppose for simplicity, that the correlation length
of $g(x,y)$ is much bigger than $l_{\omega}$.  Then it is obvious
from the above consideration that the boundary 
waves would circulate along the contours $g(x,y)=0$. If the
average value $\overline {g(x,y)}$ is negative (Fig. 5a) or
positive (Fig. 5b), and comaparable to 
$\Bigl(\overline {g^2(x,y)}\Bigr)^{1/2}$,
 these contours 
are disconnected, and, 
correspondingly,
the boundary waves are localized. As  $\overline {g(x,y)}$ approaches
zero, the contours, defined by the condition 
$g(x,y)=0$, grow in size and form a network. The correlation
size of the network divergies 
 as $\Bigl(\overline {g(x,y)}\Bigr)^{-\nu_1}$, where
$\nu_1 \approx 4/3$ is the critical exponent of the
percolation problem in two dimensions\cite{aharony}. Along with decreasing of $\overline {g(x,y)}$, the
neighboring contours $g(x,y)=0$ come closer than $l_{\omega}$, and
 the coupling of the boundary waves, encircling these contours, becomes
increasingly important (above we studied this coupling  for the
simplest case of parallel boundaries). As a result of this coupling, the interference
of different unidirectional paths comes into play. The crucial role
of interference effects, allowed by the coupling, 
was first pointed out in a seminal 
paper Ref. \onlinecite{cha-ko} in relation to
the integer quantum Hall effect. In Ref. \onlinecite{cha-ko}
unidirectional waves modeled the motion of a two-dimentional 
electron in a strong perpendicular magnetic field and a smooth
random potential (edge states). It was demonstrated in Ref. \onlinecite{cha-ko} that, with interference
taken into account, the delocalization transition in the system of
edge states occurs at {\em discrete} energies, i.e. it remains infinitely
sharp (as it is the case for the classical percolation).
However, due to the coupling and interference, the size of the eigenstates
in the critical region (localization radius) is  much bigger than the
correlation radius of the network and divergies with the exponent\cite{cha-ko}
$\nu_2\approx 2.3$. The correspondence between the edge states
of electrons and the boundary electromagnetic waves allows us
to conclude that with $\overline {g(x,y)}\rightarrow 0$ the
radius of localized boundary waves behaves as $\Bigl(\overline {g(x,y)}\Bigr)^{-\nu_2}$.

By analogy to the integer quantum Hall  transition, which originates
from the
competition of unidirectional motion of an electron 
in a magnetic field and quantum interference\cite{pru83,dima}, the
delocalization transition of the boundary waves 
at $\overline {g(x,y)} =0$ (when optical activity of the system is zero {\em on}
{\em average}) can be called quantum Cotton-Mouton effect.

Note in conclusion, that throughout the paper we assumed that the
system is uniform in the $z$-direction. The situation relevant
for experiment is a thin-film geometry, in
which two contacting optically active media  are confined
within the region $|z|<\frac{d}{2}$ with the 
 thickness, $d$, of the film being  much smaller than $l_{\omega}$. If $\varepsilon_0$
exceeds the dielectric constants of the media, between which 
the film is  sandwiched, the solutions of the Maxwell equations
are the waveguide modes, propagating along the plane of the film. 
The components of electric and magnetic fields in these modes
are  confined within
the region of the order of $d$.
Suppose first, that $g$ is constant within the film: 
$g(z)=g_0\theta(\frac{d}{2}-|z|)$. To modify the Cotton-Mouton dispersion law Eq. (\ref{cotton}) to the case of the waveguide mode propagating in the  $y$-direction, it is
convenient to rewrite the system of Maxwell's equations in the following form
\begin{equation}
\label{TE}
- \frac{\partial^2E_x}{\partial z^2} - (\varepsilon_0(z) \frac{\omega^2}{c^2} - k^2)E_x =
-\frac{\omega^2}{c^2}\frac{g^2(z)}{\varepsilon_0(z)}E_x - \frac{\omega}{c}\frac{g(z)}{\varepsilon_0(z)}\frac{\partial B_x}{\partial z},
\end{equation}
\begin{equation}
\label{TM}
- \varepsilon_0(z) \frac{\partial}{\partial z}\Biggl( \frac{1}{\varepsilon_0(z)}\frac {\partial B_x}{\partial z} \Biggr) - \Biggl(\varepsilon_0(z) \frac{\omega^2}{c^2} - k^2 \frac{\varepsilon_0(z)}{\varepsilon_1(z)}\Biggr)B_x = \frac{\omega}{c} \frac {\partial}{\partial z}\Biggl(g(z)E_x\Biggr ),
\end{equation}
where $\varepsilon_0(z)$ and $\varepsilon_1(z)$ describe the profile of the
diagonal components of $\hat \varepsilon$ in $z$-direction.
For $g_0 = 0$ the right-hand sides in Eqs. (\ref{TE},\ref{TM}) are zeros, so that the 
 above 
equations yield the sets of TE and TM waveguide modes, respectively. With the right-hand sides
included, the correction to the wave vector $k$ is quadratic in $g$. For TE mode
with a number $n$ after some algebra one can get the following dispersion law
\begin{equation}
\label{disp_}
k^2 =
(k_{TE}^{(n)}(\omega))^2 -\frac{\omega^2}{c^2}\Bigl[ (g_{eff}^{(1)})^2 - (g_{eff}^{(2)})^2
\Bigr],
\end{equation}
where
\begin{equation}
\label{bethoven1}
 (g_{eff}^{(1)})^2= g_0^2\Biggl[ \frac {\int_{-d/2}^{d/2} dz \frac {(E_x^{(n)})^2}{\varepsilon_0(z)}}{\int_{-\infty}^{\infty} dz (E_x^{(n)})^2}\Biggr],
\end{equation}
and
\begin{equation}
\label{bethoven2}
 (g_{eff}^{(2)})^2= g_0^2\Biggl[ 
\sum_{m} \frac {\int_{-\infty}^{\infty}  dz \frac {B_x^{(m)} E_x^{(n)}}{\varepsilon_0(z)}}
{\int_{-\infty}^{\infty} dz B_x^{(m)} E_x^{(n)} \frac{\partial^2}{\partial z^2}(\varepsilon_0(z)^{-1/2})} \cdot \frac{
{\int_{-d/2}^{d/2} dz \frac{ E_x^{(n)}}{\varepsilon_0(z)} \frac{\partial B_x^{(m)}}{\partial z} 
\int_{-d/2}^{d/2} dz E_x^{(n)} \frac{\partial}{\partial z}\Bigl(\frac{B_x^{(m)}}{\varepsilon_0(z)}\Bigr) }}  
{\int_{-\infty}^{\infty} dz (E_x^{(n)})^2 \int_{-\infty}^{\infty} dz \frac {(B_x^{(m)})^2}{\varepsilon_0(z)}}
                     \Biggr].
\end{equation}
The correction $(g_{eff}^{(1)})^2$ originates from the $g^2$-term in the r.h.s. of Eq. (\ref{TE}), whereas $(g_{eff}^{(2)})^2$ results from the mixing of TE mode $n$ with all TM modes.
Generally 
speaking, $g_{eff}^{(1)}$ and $g_{eff}^{(2)}$ are of the same order. This means
that, when 
$g$ depends on $x$ and changes sign, the corresponding  boundary wave would represent 
a mixture of TE and TM modes.
The situation simplifies if, for numerical reasons, $g_{eff}^{(1)}$ appears to be much bigger
than $g_{eff}^{(2)}$. In this case the dispersion law for the boundary wave, analogous to
Eq. (\ref{first}), takes the simple form $k = k_{TE}^{(n)}(\omega)$, and the electromagnetic field decays
away from the boundary $x = 0$ as $\exp \Bigl(-\frac{\omega}{c} g_{eff}^{(1)} |x| \Bigr)$.

One of the authors (M. E. R.) acknowledges useful discussions with
 H. U. Baranger, Ya.~ M.~ Blanter, A. M. Dykhne,   
and D. E. Khmelnitskii. He is also gratefull to the Aspen Center for Physics
for hospitality. The work was supported by NSF grant No. DMR-9732820.

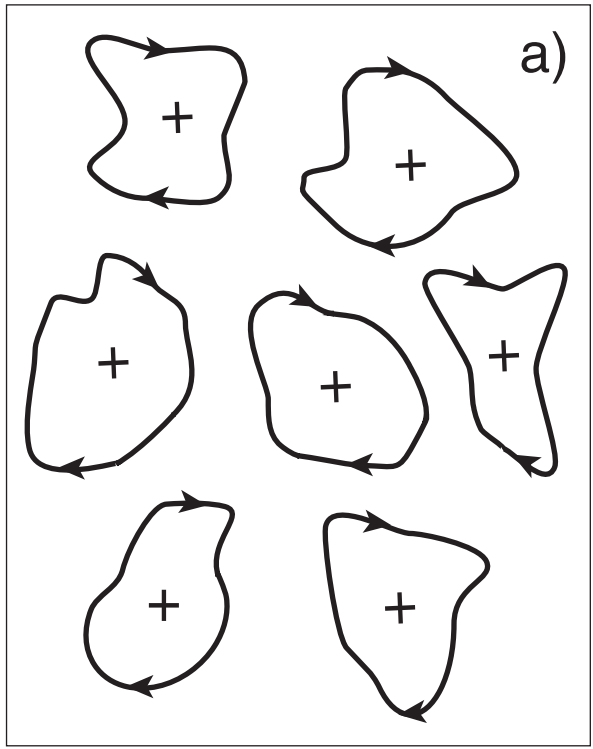
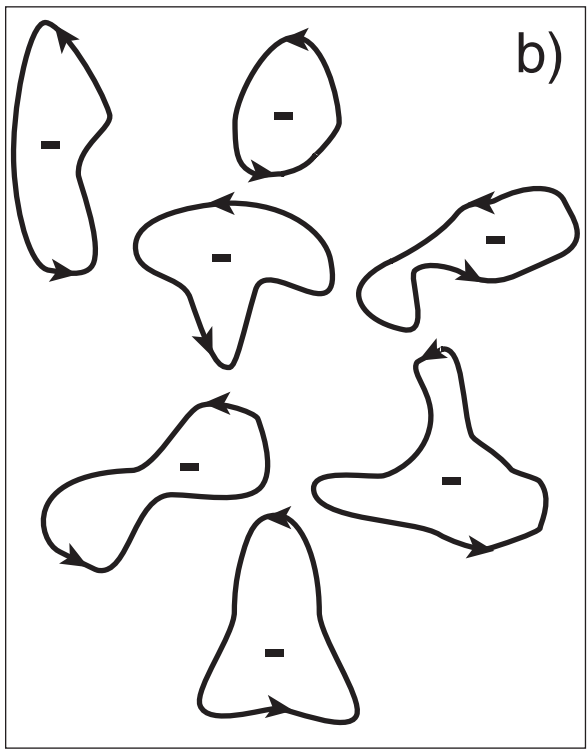
\begin{figure}
\caption{The contours $g(x,y)=0$ are shown schematically for
the case: (a) $\overline{g(x,y)} < 0$ and (b) $\overline{g(x,y)} > 0$. The arrows
show the direction of propagation of boundary waves.}
\label{fig5}
\end{figure}

\end{document}